# Janus β-$Te_2X$ (X = S, Se) Monolayers for Efficient Excitonic Solar Cells and Photocatalytic Water Splitting

Jaspreet Singh and Ashok Kumar*

*Department of Physics, Central University of Punjab, VPO Ghudda, Bathinda, 151401, India*

(December 15, 2022)

*Corresponding Author: ashokphy@cup.edu.in

## Abstract

Highly efficient, environmental friendly and renewable sources of energy are of great need today to combat with increasing energy demands and environmental pollution. In this work, we have investigated the novel 2D allotropes i.e., β-$Te_2X$ (X = S, Se) using first-principles calculations and study their potential applications in light harvesting devices. Both the monolayers possess to have the high stability and semiconducting nature with an indirect band gap. The high carrier mobilities and excellent optical absorption of these monolayers make them potential candidates for solar conversion applications. We have proposed the type-II heterojunction solar cells and calculated their power conversion efficiencies (PCEs). The small conduction band offset and appropriate band gap of donor material in case of β-$Te_2S$(S-Side)/α-$Te_2S$(Te-Side) heterojunction results in the PCE of ~ 21%. In addition to that, the band alignments of these monolayers properly engulf the redox potentials of the water. The overpotentials required to trigger the hydrogen reduction (HER) and water oxidation (OER) half reactions reveal that HER and OER preferred the acidic and neutral mediums, respectively. The calculated solar-to-hydrogen (STH) efficiencies of β-$Te_2S$ (β-$Te_2Se$) monolayers come out to be ~ 13 % (~12 %), respectively, which implies their practical applications in water splitting. Thus, our work provides strong evidence regarding the potential applications of these materials in the field of light harvesting devices.

## 1. Introduction

Clean and renewable energy sources have great relevance in today's economy and society in order to meet energy needs while lowering environmental issues. Solar energy conversion applications such as photovoltaics and photocatalytic water splitting can make wonders for the modern world. Since the discovery of solar cells in 1954,[1] continuous efforts have been made by researchers to efficiently utilize solar energy. Excitonic solar cells have entered the mainstream in recent years due to their excellent power conversion efficiencies (PCEs). These solar cells constitute of type-II van der Waals heterostrutures working on the donor-acceptor mechanism.[2, 3] The appropriate donor band gap (1.0-1.6 eV) and minimal conduction band offsets are the two main criteria for highly efficient solar cells.[4] In addition to this, high migration and low recombination of photogenerated charge carriers enhance the PCEs of excitonic solar cells. These excitonic solar cells include the organic solar cells[5-11] and inorganic two dimensional (2D) heterojunction solar cells.[12-14] In recent years since the discovery of graphene,[15] two-dimensional (2D) excitonic solar cells have attracted a lot of interest because of their high carrier mobilities, tunable band gap and good light harvesting capabilities.[12-14, 16]

Apart from solar cells, photocatalytic water splitting is also a clean and renewable source of energy that can produce clean chemical hydrogen energy by utilizing the solar energy. For an efficient photocatalyst, its solar-to-hydrogen (STH) efficiency must exceed 10% for economical commercial use.[17] In order to achieve such efficiency, the materials need to have high migration of charge carriers with low recombination, suitable band edge positions covering the redox potentials of water and high solar energy harvesting capabilities. The majority of these properties are missing in conventional three-dimensional (3D) photocatalysts leads to their low STH efficiency.[18-21] In the last decade, 2D materials have emerged as efficient photocatalysts because of their high carrier mobilities with shorter migration paths, large surface area (leads to abundant active sites) and high light harvesting abilities.[22-24] In addition to that, the tuning of their electronic properties with strain, functionalization, etc. can also be utilized to increase their STH efficiency.[25, 26] The first principles study has been widely used to predict the properties of novel 2D materials. For instance, the $PdSeO_3$ monolayer is initially predicted to have a potential use in the photocatalytic water splitting applications[27] which is experimentally synthesized later on.[28] Similar research have been conducted on various 2D materials, including GeSe,[29-31] GeTe,[32, 33] $SnS_2$,[34, 35] etc. Thus, first principles investigation of 2D materials is widely accepted and acts as a fingerprint for experimentalists. Many 2D materials are therefore being investigated experimentally or theoretically for photocatalytic water splitting, including BN,[36] phosphorene,[37] $PdSeO_3$,[27] β-$PdX_2$ (X= S, Se),[38, 39] etc.

More intriguingly, a new class of 2D materials called Janus materials has an inherent electric field caused by out-of-plane asymmetry. Such intrinsic electric field improves the spatial distribution of charge carriers by reducing their recombinations, which is advantageous for applications involving efficient solar energy conversion.[40] Additionally, these 2D Janus materials offer distinct atomic species for realizing the spontaneous hydrogen evolution reaction (HER) and oxygen evolution reaction (OER). This results in the boosting of the STH efficiency of photocatalysts such as MoSSe,[41] $B_2P_6$,[42] WSSe,[43] WSeTe,[44] $Pd_4X_3Y_3$ (X, Y = S, Se, and Te; X ≠ Y),[45] etc. Also, PCEs of heterojunction solar cells based on Janus transition metal dichalcogenides (TMDs),[46] $In_2SeTe$,[47] α-$Te_2S$,[48] etc are also found to be enhanced.

So far, mostly the solar energy conversion applications of Janus TMDs are explored. However, monoelemental 2D materials from group-VI have recently gained popularity due to their advantages over traditional 2D materials, including good stability, adequate band gap, and strong carrier mobilities.[49-52] In this work, we have explored the Janus β-$Te_2X$ (X = S, Se) monolayers and their potential applications in photovoltaic and photocatalytic water splitting. Firstly, we have analyzed the stability of these monolayers by phonon and ab-intio molecular dynamics (AIMD) simulations. Through indirect band gaps, the electronic properties have demonstrated their semiconducting nature. By creating type-II heterojunctions with other 2D materials, the solar cell applications of these monolayers are examined in the light of their semiconducting nature with suitable band gaps. Furthermore, the band alignments of these monolayers are found to straddle the redox potentials of the water. Thus, we have analyzed their ability for overall water splitting in terms of Gibbs free energy profiles in acidic and neutral mediums. Fortunately, these monolayers' STH efficiencies exceed the minimum criterion of 10% for commercial use. Finally, we examined the stacking effect on these monolayers' capacity for photocatalysis.

## 2. Computational Details

The Quantum ESPRESSO package[53] is used to perform the density functional theory (DFT)-based computations. For the treatment of electron-ion interactions and exchange-correlation energies, the projector augmented wave (PAW) potentials[54] and generalized gradient approximation (GGA) are employed, respectively. The plane wave energy cutoff of 80 Ry and k-point mesh of 16x16x1 is used for Brillouin zone integration. The structural parameters and atomic positions are relaxed up to the force and energy convergence of $10^{-6}$ eV/Å and $10^{-7}$ eV, respectively. A vacuum of 12 Å is applied to eliminate the unphysical forces between the neighboring layers along z-direction. Using the density functional perturbation theory (DFPT), the phonon dispersion is evaluated using a q-point mesh of 12x12x1 and a convergence threshold of $10^{-18}$ Ry.[55] The AIMD simulations are done using the Nose-Hoover thermostat.[56, 57] The sophisticated HSE06 hybrid functional[58] is used to evaluate the electronic properties that overcome the underestimation of band gap by Perdew-Burke-Ernzerhof (PBE) functionals.[59] In order to deal with the long range van der Waals interactions in heterostructures and bilayers, Grimme's DFT-D3 approach is used.[60] The dipole corrections have been employed to evaluate vacuum potentials because of asymmetric layer arrangements of Janus structures. To evaluate the optical properties involving the exciton effects, the YAMBO code[61] incorporated with the Quantum ESPRESSO package is used. After carrying out the DFT calculations, the $G_0W_0$ method is used to evaluate the quasi-particle energies using the plasmon-pole approximations. Then, the Bethe-Salpeter equation (BSE) is solved to evaluate the dielectric response of the system involving the electron-hole interactions. The convergence is assured by taking the total of 200 bands and k-point mesh of 20x20x1. For obtaining the optical absorbance spectra, three valence bands and three conduction bands are used to solve the Bethe-Salpeter equations. Additionally, we employed the implicit solvent model in the Environ code[62] of Quantum ESPRESSO utilizing the 78.3 dielectric constants for liquid water to account for solvent effects in simulations. The free energy and thermodynamic potential simulation details are provided in ESI.

## 3. Results and Discussion

### 3.1 Geometric Structure and Stability

Fig. 1 presents the optimized structures of Janus β-$Te_2X$ (X = S, Se) monolayers. It consists of alternatively arranged four and six-membered rings similar to that of β-Te[49] with the replacement of one surface Te atom with X (S, Se) atom. The central Te atom (marked by position 1, Fig. 1) is coordinated to four neighboring atoms, while the surface Te and X atoms are coordinated to three neighboring atoms (marked by position 2, Fig. 1). The relaxed lattice vectors of the β-$Te_2S$ ($\beta$-$Te_2Se$) rectangular unit cell are as follow: a = 4.08 Å (4.10 Å), b = 5.30 Å (5.46 Å) and the out of plane thickness 'd' is 1.82 Å (1.92 Å). These lattice parameters show slight contraction when compared to their pristine monolayer β-Te (a = 4.22 Å, b = 5.69 Å and d = 2.07 Å), which is due to the smaller size of the X atom compared to that of Te atom. The replacement of Te atom with X atom induces an intrinsic electric field from Te to X atom due to the electronegative difference of the surface atoms. This intrinsic electric field causes the potential difference (ΔV) of 0.88 eV (0.52 eV) between the Te and S (Se) sides (Fig. S1, ESI).

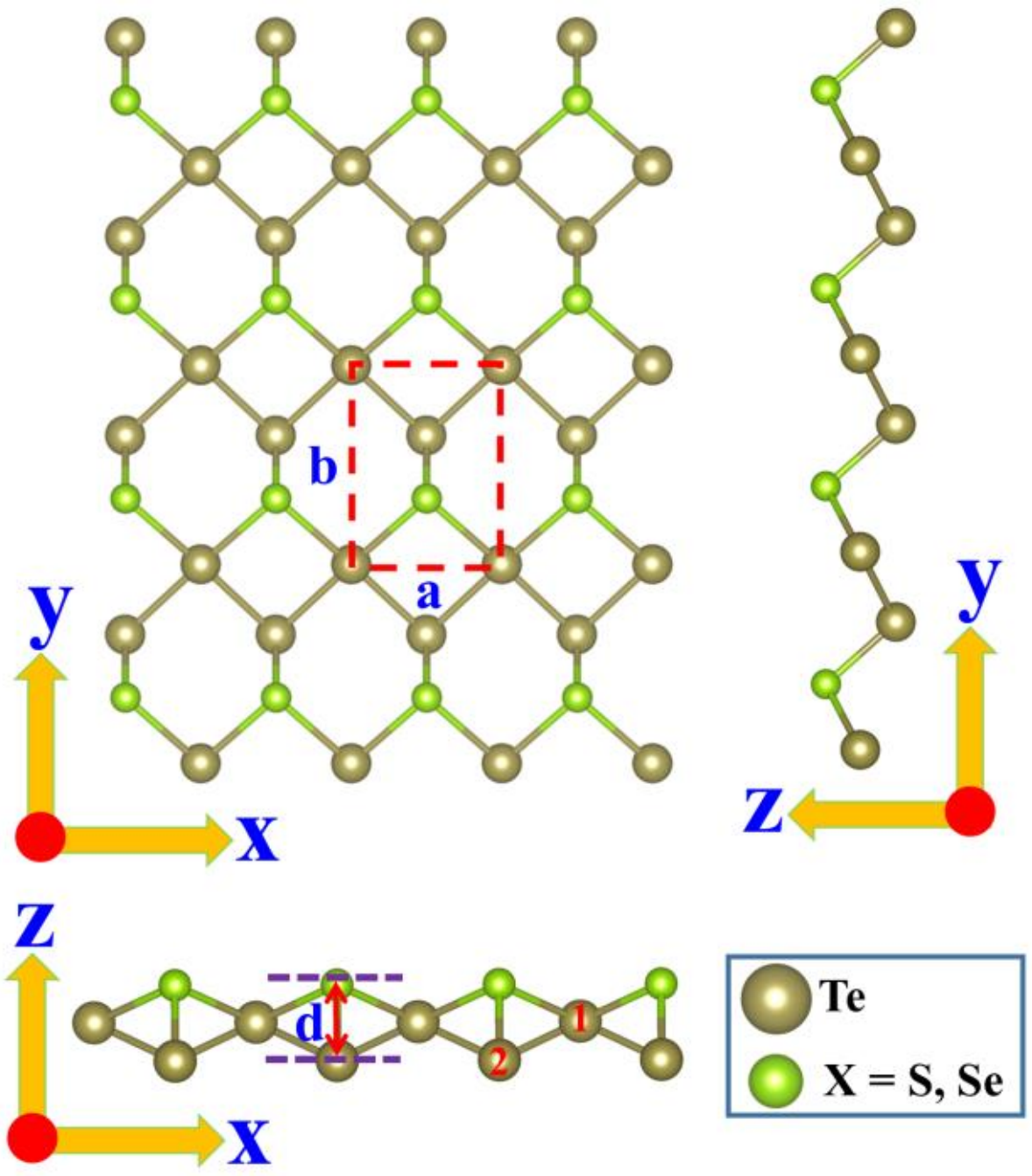

**Fig. 1** Top and side views of the crystal structure of $\beta$-$Te_2X$ (X = S, Se) monolayers.

We have now examined the stability of Janus $\beta$-$Te_2X$ (X = S, Se) monolayers. Firstly, we have calculated the cohesive energy ($E_{coh}$) of monolayers as: $E_{coh} = (2E_{Te} + E_X - E_{Te_2X})\,)/3$ ; where $E_{Te}$, $E_X$ and $E_{Te_2X}$ are the total energies of single Te, X atom and $\beta$-$Te_2X$ unit cell, respectively. The cohesive energy values of $\beta$-$Te_2S$ ($\beta$-$Te_2Se$) come out to be 3.37 eV/atom (3.21 eV/atom), which is higher than that of β-Te (3.06 eV/atom) and α-Te (3.05 eV/atom)[52] suggests that Janus $\beta$-$Te_2X$ (X = S, Se) monolayers are energetically stable.

Furthermore, we have assessed the dynamic stability of these monolayers in terms of phonon spectrum over the Brillouin zone as shown in Fig. S2(a, c), ESI. The absence of substantial imaginary phonon modes in the spectrum indicates the kinetic stability of these monolayers. The small negative frequencies in the vicinity of the Gamma (Γ) point may be considered as the soft modes which are the artifact of the computational parameters.[63, 64] The thermal stability of these monolayers is analyzed by the AIMD simulations that have been done with a relatively large supercell (4x4x1) for 5000 fs at 300K with a time step of 3 fs (Fig. S2(b, d), ESI). The smaller fluctuations of total energy and temperature around the constant level with the time and preserved crystal structures after the 5000 fs (inset of Fig. S2(b, d), ESI) suggests the thermal stability of these monolayers.

Also, the mechanical stability of these monolayers is assessed in terms of stress-strain curves (Fig. S3). The uniaxial strain is applied along the x and y-directions, corresponding to which the maximum tensile strength and strain are evaluated. The axis orthogonal to the stress direction is fully relaxed to make sure that the applied strain is uniaxial. To overcome the dimensional constraint, the equivalent stress is obtained by rescaling it with $z/d_0$, where z is the vacuum length along the z-direction and $d_0$ is the effective thickness of the system.[65, 66] The obtained values for maximum stress along x and y-directions of β-$Te_2S$ (β-$Te_2Se$) are 1.16 GPa (1.27 GPa) and 7.05 GPa (7.57 GPa), corresponding to the critical strain limits of 12% (10%) and 20% (34%), respectively (Fig. S3(a, c), ESI). All these values are smaller than that of their corresponding values for β-Te[67], especially along the x-direction. This is due to the smaller Te-X bonds as compared to that of Te-Te bonds and their tendency to transform into the 1D helical atomic chains[68, 69] when strained along the x-direction. The obtained values of Young's modulus of β-$Te_2S$ (β-$Te_2Se$) are 16.16 GPa (17.81GPa) and 37.34 GPa (30.08 GPa) along x and y-directions, respectively. The difference in Young's modulus along x and y-directions can be attributed to the anisotropic response to strain in β-$Te_2X$ (X = S, Se) monolayers (Fig. S3(a, c), ESI). This anisotropic response is due to the tetragonal crystal structure which creates a large puckered structure along the y-direction, as demonstrated in Fig 1. Additionally, it appears from the charge density analysis of β-$Te_2X$ (X = S, Se) monolayers (Fig. S4, ESI) that the charge distribution along the x and y-directions is distinct. The more localized charge distribution along the x-direction compared to the overlapping distribution along the y-direction results in the formation of anti-bonding and bonding-like states along the x and y-directions, respectively. Because bonding states are stronger than anti-bonding states, it is therefore more difficult to stretch along the y-direction than the x-direction. This causes the stress-strain curve to have a greater slope along the y-direction, which results in higher Young's modulus values and less flexibility along the y-direction than the x-direction. Also, the smaller stress of β-$Te_2S$ as compared to β-$Te_2Se$ monolayers is due to the difference in the atomic radius and electronegativity of sulfur and selenium atoms that results in different bonding strength in these monolayers. We have also calculated the Poisson ratio values from transverse-axial strain relation (Fig. S3(b, d), ESI) that come out to be 0.24 (0.31) and 0.22 (0.24) along x and y-directions, respectively. These results indicate the anisotropic and higher mechanical flexibility of Janus monolayers as compared to their pristine counterpart i.e. β-Te.

### 3.2 Optoelectronic Properties

#### 3.2.1 Electronic band structure

After the confirmation of the stability of β-$Te_2X$ monolayers, we examined their optoelectronic properties. Firstly, the electronic band structures depict the semiconducting nature of β-$Te_2S$ (β-$Te_2Se$) monolayers with an indirect band gap of 2.12 eV (2.05 eV) with hybrid HSE06 functional (Fig. 2). Note that GGA-PBE band gap of β-$Te_2S$ (β-$Te_2Se$) monolayers are calculated to be 1.49 eV (1.41 eV). The valence band maximum (VBM) is located at the Γ point, while the conduction band minimum (CBM) is located between the Y and Γ points in the Brillouin zone. The projected density of states (PDOS) indicates that the VBM is mainly contributed by the p-orbitals of surface Te and X atoms, while the CBM is contributed by a mixture of p-orbitals of all the atoms. Due to the heavy Te atom, spin-orbit coupling (SOC) effect is depicted with GGA+PBE level of theory. The SOC effect results in the slight reduction of band gaps (0.11 eV and 0.17 eV for β-$Te_2S$ and β-$Te_2Se$ monolayers, respectively) due to the splitting of the bottom of the conduction band (Fig. S5, ESI). Such a reduction of band gaps is unlikely to affect the overall results regarding the band alignments of the monolayers. Therefore, computationally intensive calculations of SOC impacts are not taken into consideration for further study.

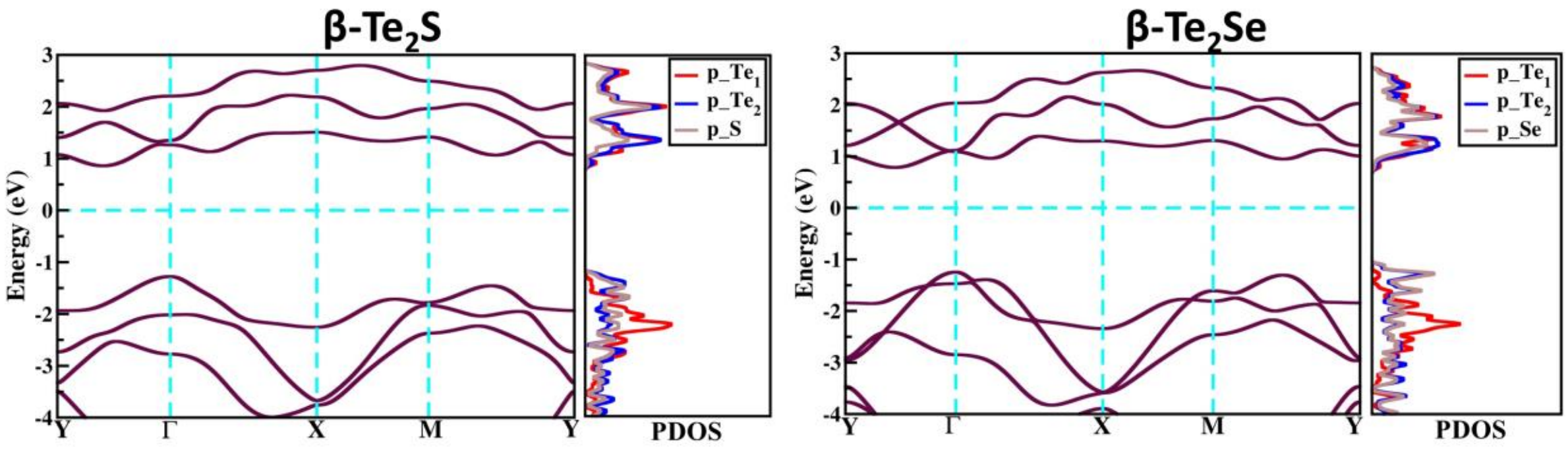

**Fig.2** Electronic band structure and PDOS of β-$Te_2X$ (X = S, Se) with HSE06 functional. The Fermi level is set at 0.

### 3.2.2 Carrier mobilities

The transportation of charge carriers is a very important aspect of materials for their applications in light harvesting devices. We have evaluated the migration of charge carriers in terms of carrier mobilities ($\mu$) using the deformation potential theory according to the longitudinal acoustic phonon limited model proposed by Lang et al.,[70] as:

$$\mu_{ax} = \frac{e\hbar^3\left(\frac{5C_x^{2D}+3C_y^{2D}}{8}\right)}{k_B T m_{ax}^{\frac{3}{2}} m_{ay}^{\frac{1}{2}}\left(\frac{9E_{ax}^2+7E_{ax}E_{ay}+4E_{ay}^2}{20}\right)} \tag{1}$$

where, $C^{2D}$, $m$ and $E$ is the elastic modulus, effective mass and deformation potential of carriers ($a$= electron or hole) and these have been assessed by fitting the curves between energy and strain (Fig. S6(a-d), ESI). Our calculated results reveal the carrier mobilities of holes (~$10^3$ $cm^2V^{-1}s^{-1}$) are one order higher than the electrons ($10^2$ $cm^2V^{-1}s^{-1}$). The values of carrier mobilities are listed in Table 1. Such higher carrier mobilities are mainly due to their smaller effective masses, and demonstrate the high migration capability of charge carriers. The difference in the mobilities of holes and electrons is primarily caused by their different effective masses, which changes their diffusion lengths ($L_p = (\mu k_b t/e)^{1/2}$) and prevents carriers from recombining.[71] The efficiency of light harvesting devices will therefore be improved by these materials' high carrier mobilities and low carrier recombination rates. These mobilities are comparable to that of β-Te[49] and higher than that of TMDs such as $MoS_2$, $MoTe_2$, $WS_2$ etc. (~$10^2$ $cm^2V^{-1}s^{-1}$).[72, 73]

**Table 1** Elastic modulus ($C^{2D}$), effective mass ($m$), deformation potential ($E$) and carrier mobility ($\mu$) of Janus β-$Te_2S$ (β-$Te_2Se$) at 300K.

| **Direction** | $\boldsymbol{C^{2D}}$ **($Jm^{-2}$)** | $\boldsymbol{m}$ **($m_0$)** | $\boldsymbol{E}$ **(eV)** | $\boldsymbol{\mu}$ **($10^3$ $cm^2V^{-1}s^{-1}$)** |
|---|---|---|---|---|
| x | 13.00 (24.55) | $m_e$= 0.49 (0.36)<br>$m_h$= 0.25 (0.17) | $E_e$= 2.96 (6.67)<br>$E_h$= 1.56 (4.41) | $\mu_e$= 0.28 (0.11)<br>$\mu_h$= 2.06 (1.15) |
| y | 17.21 (14.35) | $m_e$= 0.56 (0.67)<br>$m_h$= 0.46 (0.32) | $E_e$= 0.30 (0.86)<br>$E_h$= 0.92 (0.51) | $\mu_e$= 0.16 (0.10)<br>$\mu_h$= 1.55 (1.07) |

### 3.2.3 Optical response

Next, we have calculated the optical absorbance of β-$Te_2X$ monolayers to provide an intuitive representation of its light-harvesting capabilities. We used the $G_0W_0$+ BSE approach, which takes into account the exciton effects, to calculate the optical absorbance ($A(\omega)$) of the monolayers. We have used the following equation to evaluate the $A(\omega)$ from the imaginary part of the dielectric function ($\varepsilon_2(\omega)$) (Fig. S7 (a, c), ESI):[74, 75]

$$A(\omega) = \frac{\omega}{c} L\varepsilon_2(\omega) \quad (2)$$

where $L$ is the length of cell in the z-direction.

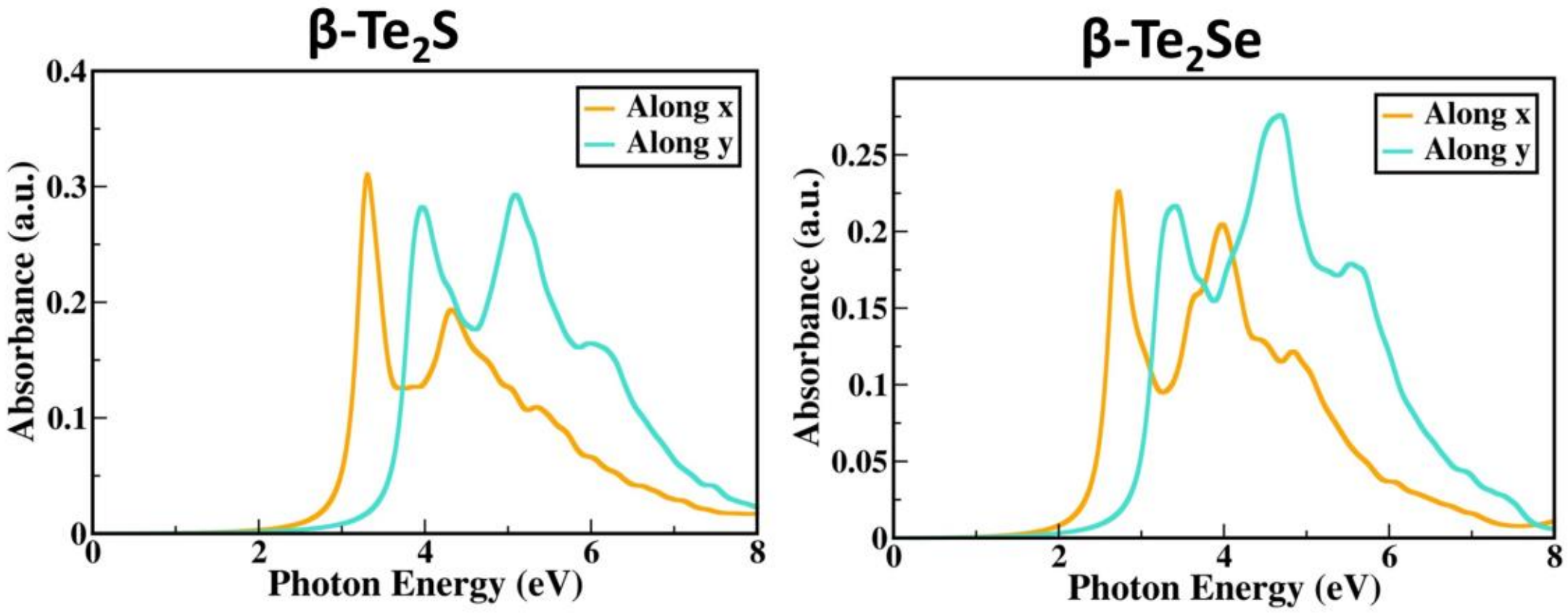

**Fig. 3** The optical absorbance of β-$Te_2X$ (X = S, Se) with $G_0W_0$+BSE method.

The absorbance spectra of both the monolayers cover most of the UV-visible region (~2-8 eV) with the first prominent peaks lying in the visible region(Fig. 3). Thus, it indicates that both the monolayers have a remarkable capacity for light harvesting. The inherent structural anisotropy of β-$Te_2X$ (X = S, Se) monolayers is responsible for the anisotropic optical absorbance. Due to this structural anisotropy, charges are localized differently along the x and y-directions (Fig. S4, ESI), which results in differing band dispersions along the x and y-directions (Fig. S7 (b, d), ESI). Different band dispersion causes the change in effective masses of electrons and holes that leads to change in e-h interactions and hence results in different optical absorbance spectra along x and y-directions. To ensure adequate free photo-excited carriers, the photon absorbed electron-hole pairs must be efficiently separated. For this, the exciton binding energy ($E_b$) is very important parameter which is calculated as: $E_b = E_{QP} - E_{OPT}$, where $E_{QP}$ is the direct quasi-particle band gap and $E_{OPT}$ is the optical gap corresponding to the first absorption peak. The quasiparticle band structures are shown in Fig. S7 (b, d), ESI. Our calculated exciton binding energies for β-$Te_2S$ (β-$Te_2Se$) monolayers are 0.48 eV (0.40 eV), which are smaller than that of β-Te (0.84 eV).[76] Due to the difference in the electronegativities of surface atoms, the built-in electric field causes the lower exciton binding energies of Janus monolayers. The more electronegative atom (X = S, Se) will attract the electrons towards itself, thus resulting in the reduction of electron-hole recombination and lowering their binding energies. As a result, these monolayers have a greater potential for extremely effective energy conversion applications.

### 3.3 Excitonic Solar Cells

Next, we have explored these monolayers for their applications in heterojunction solar cells because of their semiconducting nature, high carrier mobility and good light harvesting capability. Four type-II van der Waals heterostructures, made up of minimal lattice-mismatching β-$Te_2X$ (X = S, Se), α-$Te_2S$, and H-$MoX_2$ (X = S, Se) monolayers, have been proposed (Fig. S8, ESI). The lattice parameters, band gaps and band dispersions of constituent monolayers are given in ESI (Table S1 and Fig. S9). The low binding energies and larger interlayer distance demonstrate the van der Waals interactions between the constituent monolayers (Table S2). Therefore, the intrinsic optoelectronic properties of individual monolayers will remain preserved in the heterojunctions. Thus, we have used the Anderson method[77] to evaluate the conduction band offsets ($\Delta E_c$) by utilizing the band alignments of monolayers constituting the heterostructures.[78, 79] The different surface atoms result in the electrostatic potential difference and will induce band bending. As a result, we examined the band edge positions in relation to the vacuum levels on different surfaces that make up S, Se and Te atoms (Fig. 4(a-d)). The smaller $\Delta E_c$ and appropriate band gap of donor material will result in highly efficient solar cells. Fortunately, our proposed heterojunction (β-$Te_2S$(S-Side)/α-$Te_2S$(Te-Side) has very small $\Delta E_c$ (0.08 eV) and appropriate donor band gap of 1.49 eV that will result in highly efficient heterojunction solar cell. Note that in heterojunction solar cells, the materials that have a greater (lower) CBM value compared to the CBM of the other materials in the type-II heterostructures serve as donors (acceptors).

The PCEs of our proposed heterojunction are calculated by the widely used method proposed by Scharber *et al.*[80] At 100% external quantum efficiency, PCE's highest limit can be formulated as follows:

$$\eta = \frac{\beta_{ff} V_{oc} J_{sc}}{P_{solar}} = \frac{0.65(E_g - \Delta E_c - 0.3) \int_{E_g^d}^{\infty} \frac{P(\hbar\omega)}{\hbar\omega} d(\hbar\omega)}{\int_0^{\infty} P(\hbar\omega) d(\hbar\omega)} \tag{3}$$

where $\beta_{ff}, V_{oc}, J_{sc}$ and $P_{solar}$ are the band fill factor, open circuit voltage, short-circuit current and total incident solar energy, respectively.

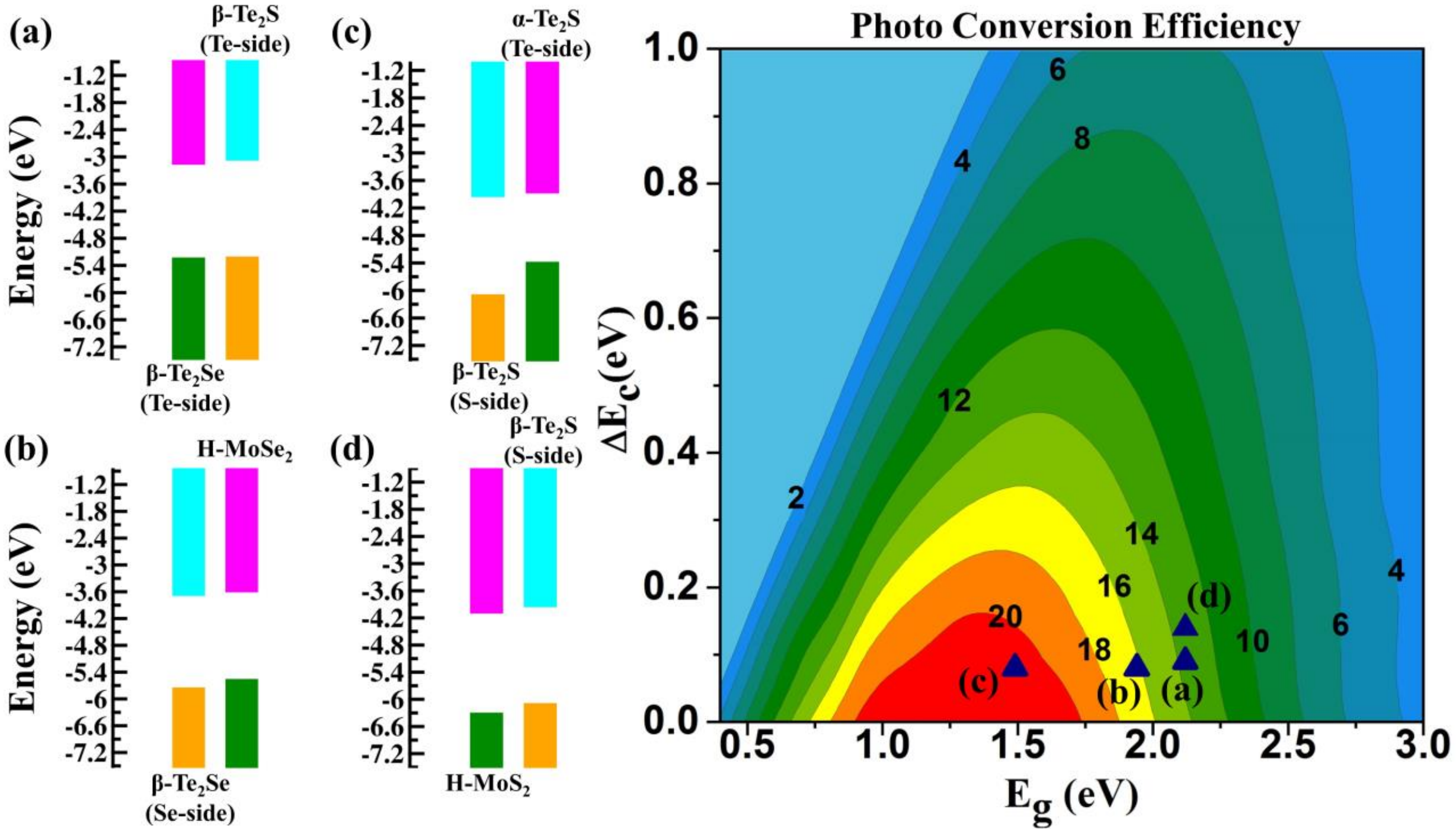

**Fig. 4** Band alignments of **(a)** β-$Te_2Se$(Te-Side)/β-$Te_2S$(Te-Side) **(b)** β-$Te_2Se$(Se-Side)/H-$MoSe_2$ **(c)** β-$Te_2S$(S-Side)/α-$Te_2S$(Te-Side) **(d)** β-$Te_2S$(S-Side)/ H-$MoS_2$ and their PCEs corresponds to the donor band gap and conduction band offset.

**Table 2** PCEs of our and previously reported 2D heterojunction solar cells.

| Heterojunctions | PCE (%) | Ref. |
|---|---|---|
| β-$Te_2S$(S-Side)/α-$Te_2S$(Te-Side), β-$Te_2Se$(Se-Side)/H-$MoSe_2$ | 21.13, 16.18 | This work |
| Te/$WTe_2$, Te/ $MoTe_2$ | 22.5, 20.1 | 78 |
| GeSe/SnSe | 21.47 | 30 |
| 2D/3D hybrid perovskites | 21.49 | 81 |
| α-$Te_2S$ (S-side)/ T-$PdS_2$, α-$Te_2S$ (Te-side)/ BP | 21.14, 19.04 | 48 |
| $HfTeSe_4$/$Bi_2WO_6$ | 20.8 | 82 |
| $MoSe_2$/Ψ-phosphorene | 20.26 | 83 |
| $Pb_2SSe$/SnSe, $Pb_2SSe$/GeSe | 20.02, 19.28 | 84 |
| $MoS_2$/bilayer phosphorene | 16-18 | 14 |
| P-$PdSe_2$/$MoTe_2$,T-$PdSe_2$/$MoS_2$(3L) | 17, 17 | 16 |
| CBN/PCBN | 10-20 | 12 |
| g-SiC2/GaN bilayer | 12-20 | 85 |

All these values are evaluated with the help of the above formulism using the donor band gap ($E_g$), $\Delta E_c$ and air mass 1.5 (AM1.5) solar energy flux at energy $\hbar w$ ($P(\hbar w)$). The $\beta_{ff}$ is approximated to be 0.65 using the Shockley−Queisser limit.[2] The β-$Te_2S$(S-Side)/α-$Te_2S$(Te-Side) heterojunction has the maximum PCE (21.13%) from our proposed heterostructures attributed to its outstanding open circuit voltage (1.11 V), lower $\Delta E_c$ and appropriate donor band gap. The PCE of our proposed heterojunctions are depicted in Fig. 4 as a function of $\Delta E_c$ and $E_g$ and their values corresponding to $V_{oc}$ are listed in Table S2, ESI. The PCEs of other heterojunction solar cells are listed in Table 2 for comparison. The PCEs of the suggested heterojunctions are comparable to or higher than many of the previously reported heterojunction solar cells.

### 3.4 Photocatalytic Properties of β-$Te_2X$

Next, we investigated the band alignments with respect to vacuum levels to unveil the photocatalytic properties of β-$Te_2X$ monolayers. In general, the material must have band gap in between 1.23 eV and 3 eV provided that its VBM and CBM must span the water oxidation (-5.67 eV at pH = 0) and reduction (-4.44 eV at pH = 0) potentials, respectively to act as water splitting photocatalyst. However, because the different surface sides of Janus materials offer different vacuum levels, they work in concert to enable the HER and OER. In our case, the conduction band edge of the Te-side and the valence band edge of the X-sides both exceed the redox potentials, providing sufficient activity for HER and OER, respectively (Fig. 5a). The external potentials of photogenerated carriers that directly impact the photocatalytic activity are calculated according to previous literature.[27, 43] The potential of photogenerated electrons ($U_e$) for HER (difference between conduction band edge and hydrogen reduction potential) and holes ($U_h$) for OER (difference between conduction band edge and hydrogen reduction potential) at pH = 0 are calculated to be 1.40 eV (1.31 eV) and 1.60 eV (1.26 eV), respectively for β-$Te_2S$ (β-$Te_2Se$) monolayers. The reduction potential will alter as a function of pH as; $-4.44 + 0.059 \times$ pH (eV). As a result, an increase in pH will cause the potential of photogenerated holes (electrons) to rise (fall). At pH = 7, the values of $U_e$ and $U_h$ for β-$Te_2S$ (β-$Te_2Se$) monolayers come out to be 0.99 eV (0.90 eV) and 2.01 eV (1.67 eV), respectively. The positive values of $U_e$ and $U_h$ are an indicator of the possible photocatalytic activity of these monolayers in neutral and acidic environments.

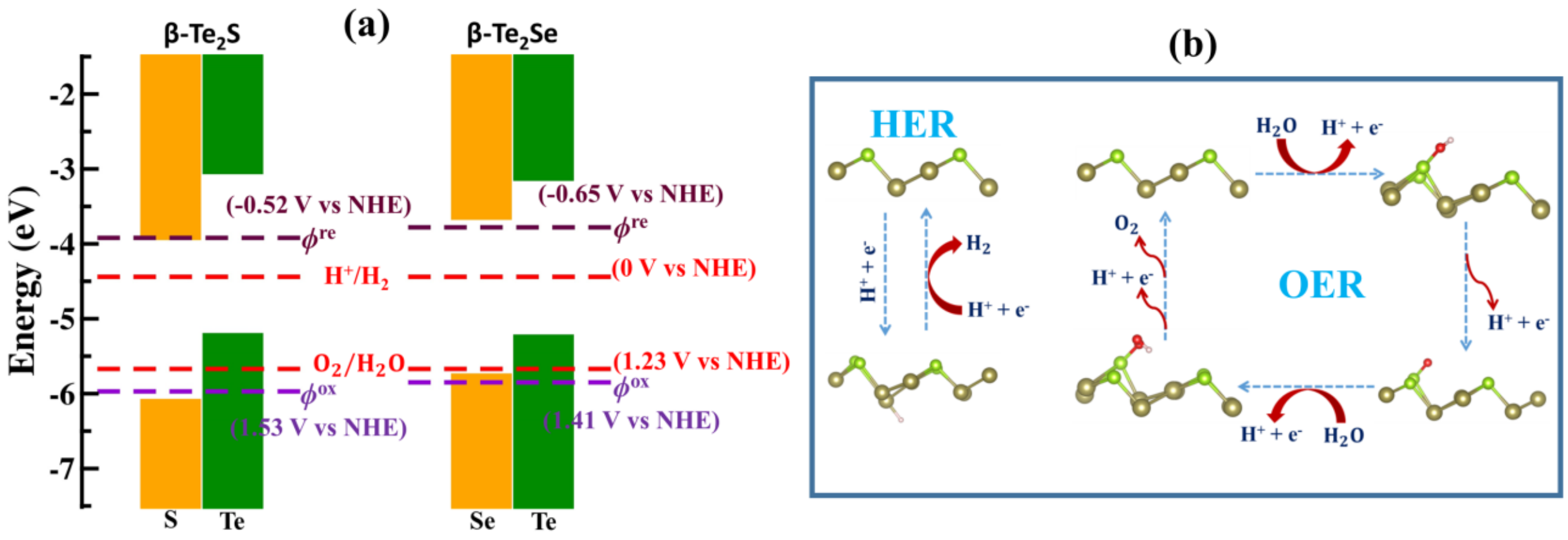

**Fig. 5(a)** Band alignments of β-$Te_2X$ (X = S, Se) monolayers at HSE06 level of theory. The vacuum level is set at zero. The red dashed horizontal lines represent the redox potentials of water at pH = 0. The brown and violet color dashed lines represent the $\phi^{ox}$ and $\phi^{re}$, respectively. **(b)** Proposed photocatalytic pathways and the atomic configuration of absorbed intermediate species for hydrogen reduction (HER) and water oxidation (OER). (Note: white balls = hydrogen atoms; red balls = Oxygen atoms).

Further, we assess the stability of photocatalyst in the aqueous solution under the illumination using the method proposed by Chen *et al.*[86] We have calculated the thermodynamic oxidation ($\phi^{ox}$) and reduction ($\phi^{re}$) potentials shown by violet and red lines in Fig. 5(a). The calculation details can be found in ESI. For both the monolayers, the $\phi^{ox}$ ($\phi^{re}$) is lower (higher) than that of the oxidation potential of $O_2/H_2O$ (reduction potential of $H^+/H_2$). This implies that the photogenerated carriers will reduce or oxidize the water molecules rather than the photocatalyst.[86] As a result, the β-$Te_2X$ monolayers exhibit good photoinduced corrosion resistance. In addition to the suitable band edge positions and stability in an aqueous solution of β-$Te_2X$ monolayers, the feasibility of adsorption of water molecules on the surfaces of these monolayers also needs to be examined. This has been done by evaluating the adsorption energies of water molecules using a model of one $H_2O$ molecule adsorbed on 2x2 supercell of β-$Te_2X$. The adsorption energy ($E_{ads}$) is calculated as:

$$E_{ads} = E_{*H_2O} - E_* - E_{H_2O} \tag{4}$$

where $E_{*H_2O}$, $E_*$ and $E_{H_2O}$ are the total energy of adsorbed monolayer with adsorbed $H_2O$ molecule, a pristine monolayer, and a water molecule, respectively. Fig. S10, ESI shows the $H_2O$ adsorption structures on both (Te and X) surfaces. The adsorption energies on the Te and X sides of β-$Te_2X$ are calculated to be approximately -0.15 eV and -0.10 eV, respectively. The negative adsorption energies indicate the feasibility of water adsorption on β-$Te_2X$ monolayers.

### 3.4.1 Overall water splitting of β-$Te_2X$ monolayers

In addition to the suitable band edge positions, the photogenerated electrons and holes in the β-$Te_2X$ monolayers need to have a sufficient driving force to initiate the overall water splitting. To this purpose, in order to acquire a better understanding of the photocatalytic activity of β-$Te_2X$ monolayers, we methodically explored the mechanisms and Gibbs free energy profiles of both the half-reactions of hydrogen reduction and water oxidation. The details of the computations are given in ESI. Note that in order to characterize the thermodynamics of chemical processes at the solid/liquid interfaces, we also incorporate the solvation effects.

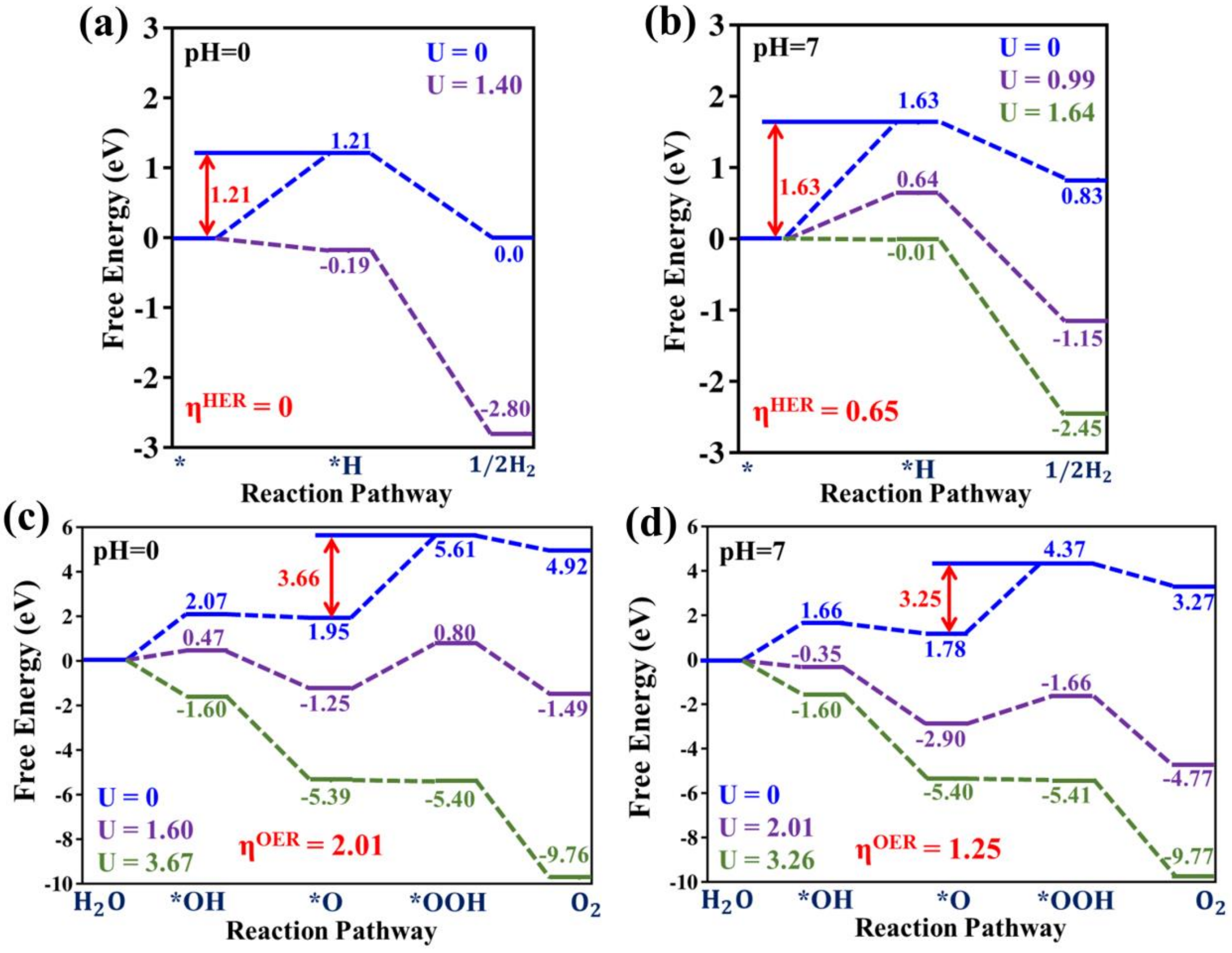

**Fig. 6** The free-energy changes for **(a, b)** HER and **(c, d)** OER at pH = 0, 7 of β–$Te_2S$ monolayer. The blue, violet and green lines represent the situation without any light irradiation, with light irradiation (at potential of photogenerated carriers) and with external potentials, respectively.

Firstly, we will discuss the much simpler hydrogen reduction (HER) on the Te-side of the β-$Te_2X$ monolayer, which is a two-step process as shown in Fig. 5(b). The first step involves the combination of proton and electron with β-$Te_2X$ monolayer to form the ∗H species; the second step is accompanied by the release of $H_2$ molecule when ∗H species bond with proton and electron. The first step is endothermic (at pH = 0, 3, 7, U = 0) in nature with positive values of free energy change (HER barrier) while the second step is exothermic in nature. Thus, the first step is not energetical favorable in the absence of light irradiation (U = 0). However, under the potential of photogenerated electrons (U = 1.40 eV and 1.31 eV for β-$Te_2S$ and β-$Te_2Se$, respectively), both the steps at pH = 0 becomes downhill, makes the HER feasible in the acidic medium under illumination (Fig. 6(a) and Fig. 7(a)). At pH = 7, external potential ($\eta^{HER}$) of 0.65 eV (0.73 eV) for β-$Te_2S$ (β-$Te_2Se$) is required to trigger the HER (Fig. 6(b) and Fig. 7(b)) and are less than that of $SiP_2$ (0.83 eV),[87] PE-$AgBiP_2Se_6$ (1.06 eV),[88] and FE-$AgBiP_2Se_6$ (1.62 eV).[88] This is due to the fact that an increase in pH lowers the potential of photogenerated electrons and raises the HER barrier. It can be concluded that the HER is more feasible on the Te-side of β-$Te_2X$ in the acidic medium than in the neutral medium.

The half reaction of water oxidation (OER) is a four-step reaction that takes place on the X-side of β-$Te_2X$ monolayers as shown in Fig. 5(b). In the first step, the oxidation of water molecule takes place to form ∗OH species; in the second step, ∗OH species oxidized to ∗O species with the release of proton and electron: the third step is accompanied by the oxidation of ∗O species with another water molecule to form ∗OOH species; finally the free oxygen molecule is released by the oxidation of ∗OOH species. The corresponding free energy profiles at pH = 0, 7 are given in Fig. 6(c-d) and Fig. 7(c-d). The step involves in the formation of ∗OOH species (third step) at pH = 0, 3, 7, U = 0 have the maximum free energy change that can be considered as an OER barrier. In the presence of potential of photogenerated holes, the third step remains uphill at pH = 0, 3, 7. Thus, to make all the steps exothermic and oxidation half reaction feasible on X-side of β-$Te_2X$ monolayers, the external potential is required ($\eta^{OER}$). With the decrease in the OER barrier and increase in the potential of photogenerated holes with an increase in pH, the required external potential to trigger the oxidation half reaction decreases. The values of $\eta^{OER}$ lower down from 2.01 eV (1.30 eV) to 1.25 eV (0.48 eV) for β-$Te_2S$ (β-$Te_2Se$) with the change of pH from 0 to 7. The values of $\eta^{OER}$ in neutral medium are lower than that of many reported 2D materials such as GaAs (1.39 eV),[89] g-C3N4 (1.45 eV),[90] CuCl (1.53 eV),[91] etc. This implies that the neutral medium is more favorable than acidic medium for OER. Note that, the free energy profiles of β-$Te_2X$ monolayer for HER and OER at pH = 3 are also provided in ESI (Fig. S11 and Fig. S12). Note that the external potentials required to initiate the HER and OER can be reduced for efficient photocatalytic activity by using several strategies, such as single-atom catalysts,[92] defect engineering,[93] strain engineering,[94] etc.

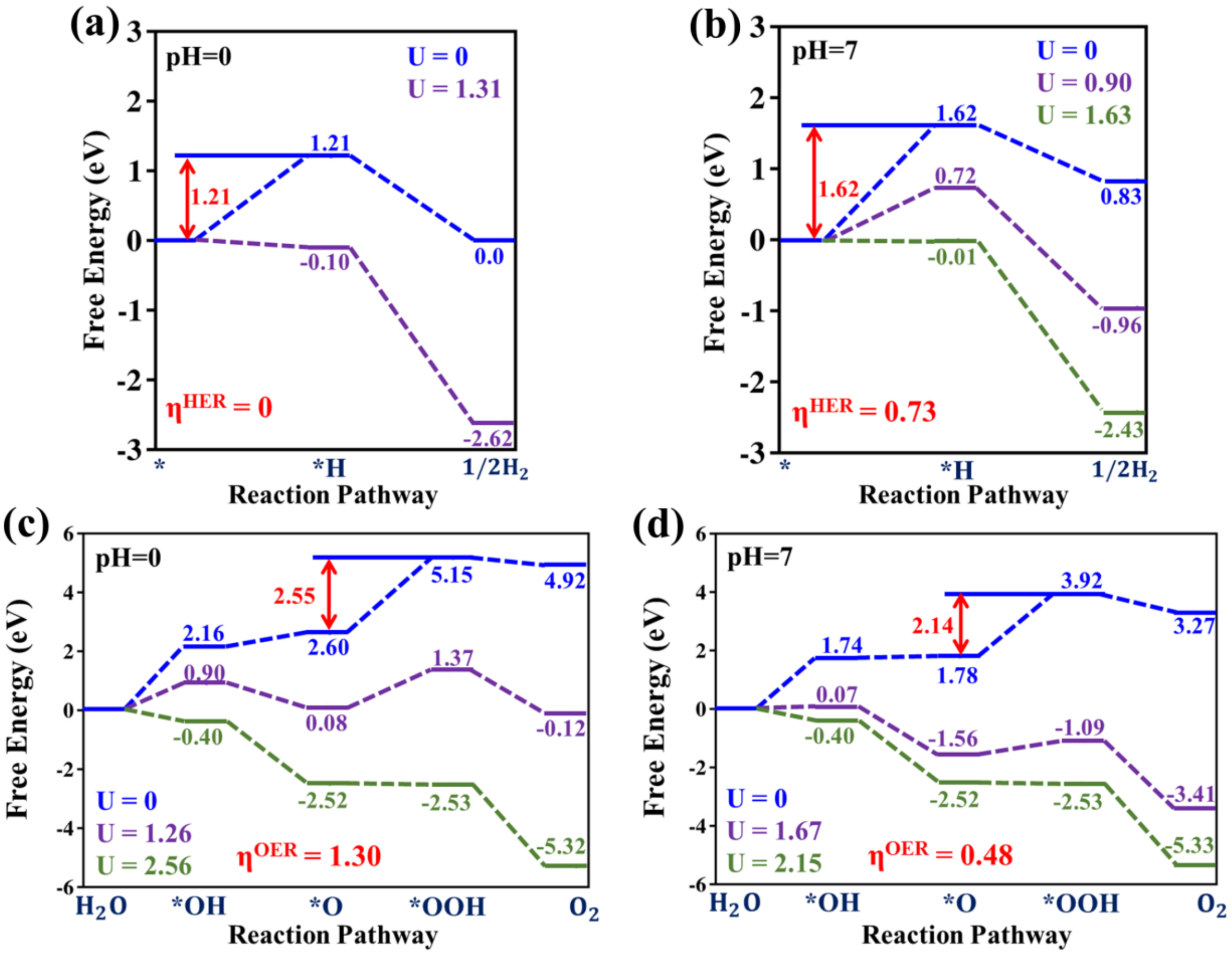

**Fig. 7** The free-energy changes for **(a, b)** HER and **(c, d)** OER at pH = 0, 7 of β–$Te_2Se$ monolayer. The blue, violet and green lines represent the situation without any light irradiation, with light irradiation (at potential of photogenerated carriers) and with external potentials, respectively.

### 3.4.2 Solar-to-hydrogen (STH) efficiency

In β-$Te_2X$ monolayers, the active sites for HER and OER are on two different surfaces, which points to these monolayers as potential candidates for efficient photocatalytic characteristics. The STH efficiency ($\eta_{STH}$) is a standard tool to measure the performance of material as a photocatalyst that depends upon the band gap and overpotentials for HER ($\chi(H_2)$) and OER ($\chi(O_2)$). The $\eta_{STH}$ is calculated as: $\eta_{abs} \times \eta_{cu}$, where $\eta_{abs}$ and $\eta_{cu}$ are light absorption and carrier utilization efficiencies, respectively. The $\eta_{abs}$ mainly depends upon the band gap values and for higher $\eta_{cu}$ the $\chi(H_2)$and $\chi(O_2)$ should have the appropriate levels. The calculation details of $\eta_{abs}$ and $\eta_{cu}$ are given in ESI. In Janus materials, the intrinsic electric field that facilitates the electron hole migration is also taken into the account in terms of potential difference ($\Delta V$) and the corrected STH efficiency ($\eta'_{STH}$)[40] is given by:

$$\eta'_{STH} = \eta_{STH} \times \frac{\int_0^{\infty} P(\hbar\omega)d(\hbar\omega)}{\int_0^{\infty} P(\hbar\omega)d(\hbar\omega) \times \Delta V \int_{E_g}^{\infty} \frac{P(\hbar\omega)}{\hbar\omega} d(\hbar\omega)} \tag{5}$$

The corrected STH efficiencies at different pH values are listed in Table S5, ESI as a function of $\chi(H_2)$ and $\chi(O_2)$. The light absorption efficiencies of β-$Te_2X$ monolayers are higher than 30% and carrier utilization efficiencies of β-$Te_2S$ (β-$Te_2Se$) monolayers reach up to 47.29% (37.58%) at pH = 7. As a consequence, the maximum corrected STH efficiencies of β-$Te_2S$ (β-$Te_2Se$) monolayers at pH = 7 are 13.46% (12.09%), respectively. These values fulfill the 10% efficiency criteria for producing hydrogen for commercial use by photocatalytic water splitting.[17] Additionally, we compared our computed STH efficiencies with previously reported values, and the results demonstrate that they are on par with or even higher than those of many reported 2D materials. (Table S6, ESI).

### 3.5 Photocatalytic Ability of β-$Te_2X$ Bilayers

Further, we have proposed different stacking patterns of β-$Te_2X$ bilayers and analyzed the photocatalytic ability of the most stable configurations in terms of their band alignments. The six kinds of stacking models denoted by AAn and ABn (n = 1, 2, 3) are depicted in Fig. S13, ESI. Firstly, we have analyzed the stability of these bilayers by calculating their binding energies ($E_b$) as:

$$E_b = 2E_{monolayer} - E_{bilayer} \tag{6}$$

where $E_{monolayer}$ and $E_{bilayer}$ are the total energies of monolayer and bilayer, respectively. The values of binding energies and their interlayer distances are listed in Table S7, ESI. The values of binding energies imply that the AB stacking models are more stable than that AA stacking models. We can divide these six stacking patterns into two types depending upon the electrostatic potential difference ($\Delta V$): AA1 and AB1; AA2, AA3, AB2 and AB3 patterns. The former type has almost double the $\Delta V$ as that of corresponding monolayers while the latter type has zero $\Delta V$ because of the healing of out-of-plane symmetry breaking of monolayers (Fig. S14 and Table S7, ESI). Out of these two types, we have chosen AB1 and AB3 stacking patterns depending upon their binding energies and electrostatic potential differences.

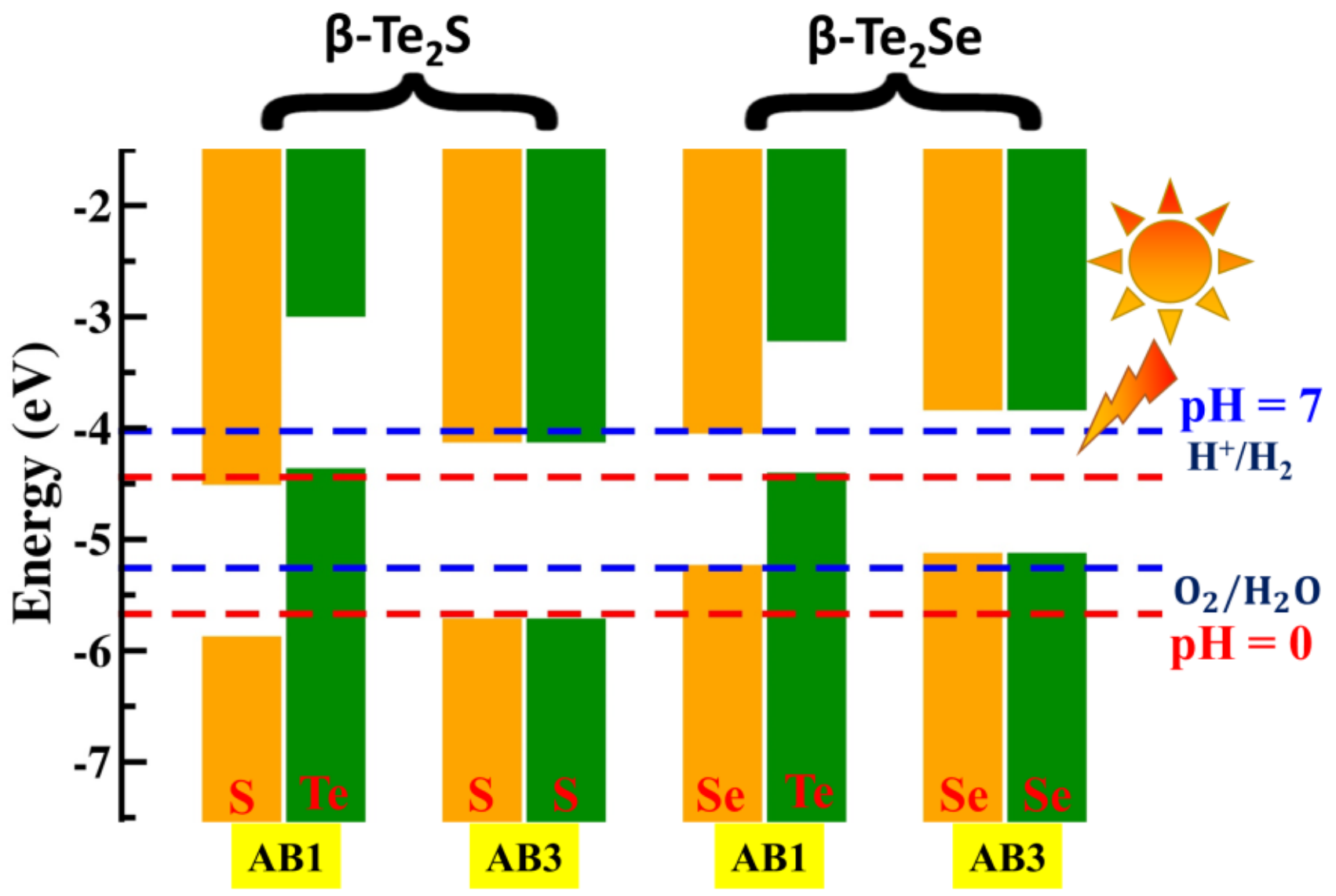

**Fig. 8** Band alignments of bilayers of β-$Te_2X$ (X = S, Se). The vacuum level is set at zero. The red and blue dashed horizontal lines represents the redox potentials of water at pH = 0 and 7, respectively.

To analyze the photocatalytic ability of these bilayers, we have studied their electronic properties using HSE06 functional. The electronic band gaps of AB1 and AB3 stacking patterns come out to be 1.36 eV (1.18 eV) and 1.58 eV (1.28 eV) of β-$Te_2S$ (β-$Te_2Se$) bilayers, respectively (Fig. S15, ESI). The band alignments with respect to vacuum levels of β-$Te_2X$ bilayers are shown in Fig. 8, which demonstrates that only the AB1 stacking of β-$Te_2S$ has proper band edge positions for photocatalytic activity over a pH range of 0 to 7, with Te- and S-sides suitable for HER and OER. However, the AB3 stacking of β-$Te_2S$ can show the photocatalytic properties in the acidic medium, while the bilayers of β-$Te_2Se$ don't have the proper band alignments for photocatalytic properties.

## 4. Conclusions

In summary, a comprehensive study based on first-principles calculations, on 2D β-$Te_2X$ (X = S, Se) monolayers has been performed for their potential applications in photovoltaic and photocatalytic water splitting applications. Firstly, these monolayers are found to exhibit energetical, dynamical, thermal and mechanical stability. The optoelectronic properties revealed their semiconducting nature with high carrier mobilities and pronounced light absorption abilities. The proposed heterojunction solar cells based on these monolayers exhibit high PCEs, especially in case of β-$Te_2S$(S-Side)/α-$Te_2S$(Te-Side) heterojunction in which the PCE reaches up to 21.13%. Further, we have investigated their photocatalytic properties as the redox potentials of water are found to be lies within the band edge positions relative to different surface vacuum potentials. The Gibbs free energy profiles reveal that the HER can proceed spontaneously in the acidic medium under illumination. However, the external potentials of 0.65 eV and 0.73 eV are required in cases of β-$Te_2S$ and β-$Te_2Se$, respectively to trigger the HER in the neutral medium. In contrast to the HER, the OER is more favorable in a neutral medium with required external potentials of 1.25 eV and 0.48 eV for β-$Te_2S$ and β-$Te_2Se$, respectively. Greater than 10% STH efficiency of these monolayers in the neutral environment describes their potential practical applications in the commercial production of hydrogen. At last, the stacking impact on these monolayers' photocatalytic activity shows that only bilayers of β-$Te_2S$ have the proper band alignments for photocatalytic water splitting. Thus, these monolayers can be the potential candidate for futuristic solar energy conversion devices.

## Acknowledgments

JS is thankful to the Council of Scientific and Industrial Research (CSIR) for providing financial assistance in the form of the Senior Research Fellowship (SRF). The computational facility available at the Central University of Punjab, Bathinda, was used to obtain the results presented in this paper. We also like to acknowledge Mukesh Jakhar for his helpful discussions.